\documentclass{iau} 
\usepackage{graphicx}

\title[LSST and the Milky Way] 
{The Large Synoptic Survey Telescope and Milky Way Science}

\author[R. Michael Rich]   
{R. Michael Rich $^1$}

\affiliation{$^1$ Dept. of Physics and Astronomy, UCLA, Los Angeles, CA 90095-1547 \\ email: {\tt rmr@astro.ucla.edu}} 
\pubyear{2017}
\volume{334}  
\jname{Rediscovering our Galaxy}
\editors{C. Chiappini, I. Minchev, E. Starkenburg, \& M. Valentini, eds.}
 
\begin{document}

\maketitle

\begin{abstract}
The Large Synoptic Survey Telescope (LSST) surveys have initially been optimized to omit the inner part of the Milky Way disk/bar from deep and cadence observations.  However it is now clear that the LSST will be powerful for Galactic astronomy and may play a crucial role in continuing to extend the Gaia astrometric catalog until a future satellite, either optical or IR, carries on.   LSST will provide metallicities and kinematics for the bulge, and will map halo structures to as distant as 450 kpc, nearly half the distance to the Andromeda galaxy.   Thanks to the unprecedented calibration effort for its photometric system, and surprisingly good astrometry (transverse velocity measurements of 0.2 mas/yr at r=21; 1 mas/yr at r=24) {\it LSST will provide photometric abundances and distance constraints for a billion or more Milky Way stars to distances of 450 kpc, and kinematics from proper motions to $\sim100$ kpc}.   Single observation depths reach $\sim 24$ in the $ugrizy$ bands, while depths at end of mission reach $\sim 27$.   Although halo structures such as streams and dwarf galaxies are initially identified by the RR Lyrae and giants, their structure will be fleshed out by the 100$\times$ more abundant dwarfs that will be detected to 100 kpc (single observation) and  $\sim 300$ kpc by end of mission.  More complete mapping of stream structures may constrain the mass distribution of dark matter and perhaps confirm the interaction of dark matter halos and streams.   I also describe the Blanco DECam Bulge Survey, a 200 deg$^2$ LSST pathfinder survey of the bulge in $ugrizy$ using the Dark Energy Camera on the Blanco 4m telescope.   The purpose of this article is to encourage active workers on the Milky Way and Local Volume to participate in the LSST project, in particular to urge that the Galactic Plane receive the same cadence and depth coverage as the rest of the extragalactic sky.

\keywords{telescopes, survey, Galaxy: structure, Galaxy: bulge, Galaxy: disk, Galaxy: halo}
\end{abstract}

\section{Introduction}

\begin{figure}
\begin{center}
 \includegraphics[width=3.0 in]{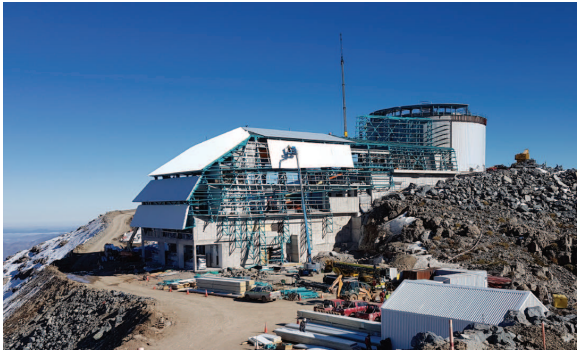} 
 \caption{Construction progress of LSST at Cerro Pachon, Chile as of 14 July 2017;  for current progress see \textit {https://gallery.lsst.org/bp/\#/}  }
  \label{fig1}
\end{center}
\end{figure}

If one contemplates the core science goals of LSST, an extragalactic or pure transient science case is first to come to mind.  Indeed, the four core science themes are:  Dark energy and dark matter (via measurements of strong and weak lensing, large-scale structure, clusters of galaxies, and supernovae), exploring the transient and variable universe, structure of the Milky Way galaxy and its neighbors via resolved stellar populations, and constructing an inventory of the Solar System, including everything from near Earth asteroids to the Kuiper Belt.  The Milky Way science case is included, but the cosmology and transient science cases take highest priority.   And it is important to remember that the mission of detecting potentially hazardous asteroids helped secure funding for LSST and has its importance (ask any dinosaur).  So the Milky Way is (on paper at least) a lower priority for LSST; in fact, LSST stands to have a significant impact on Milky Way science.  Gaia and LSST have a largely unappreciated but crucial synergy.  Gaia will help to improve the astrometric quality of LSST, and LSST will enable extension of Gaia data to fainter magnitudes and potentially, may effectively help to extend the Gaia mission by updating its astrometric catalog.  LSST has an additional remarkable feature: photometric stability assured by the use of a dedicated photometric monitor telescope.  The quality of LSST photometry means that one must take seriously the possibility that the 6 color photometry of $ugrizy$ may give photometric abundances and parallaxes of unprecedented quality, enabling studies along the lines of Ivezi\'{c} et al. (2008) to be carried to 400 kpc from the Milky Way.    The reader interested in a deeper exploration should be aware of the LSST Science book (LSST Collab. et al. 2012), the Ann. Rev. article by Ivezi\'{c}, Beers, \& Juri\'{c} (2012), and the Science-Driven optimization of the LSST Observing Strategy (Marshall et al. 2017); this reference is crucial for those proposing new observations.  Those interested in the Gaia-LSST synergy and astrometry with LSST may consult Bernstein et al. (2017a,b) and Pier et al. (2003)  that discuss astrometric calibration and performance, and the instrumental response model for the Dark Energy Camera and the astrometric solution for SDSS respectively; the LSST focal plane will be similar to that of the DECam at the prime focus of the 4m telescope.   The Milky Way and Local Volume Collaboration has this subject area as its focus; present co-chairs are J. Gizis, N. Kallivayalil, and J. Bochanski.  Figure \ref{fig1} shows the status of LSST as of 14 July 17; the project is real, and the reader can visit the same website to see construction progress on any date.

\begin{figure}
\begin{center}
 \includegraphics[width=3.0 in]{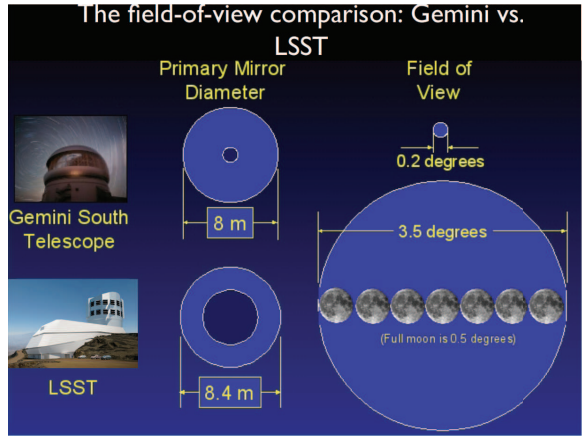} 
 \caption{LSST compared to Gemini.  The effective area of the telescope mirror (6.7m) is smaller, but the field of view
 is substantially larger than even the hypersuprime camera at Subaru.  This gives a visual graphic of the factor of 10 greater \'etendue of LSST (m$^2$ of collecting area $\times$ field of view in deg$^2$).  The LSST mirror is an 8.5m monolith, but the central 5m is a tertiary mirror, the last reflection into the camera (the second reflection is a 3.4 m convex secondary). The optics enable the huge field of view (source: Chuck Claver/ LSST project).
 }
   \label{fig2}
\end{center}
\end{figure}

\begin{figure}
\begin{center}
\includegraphics[width=3.0 in]{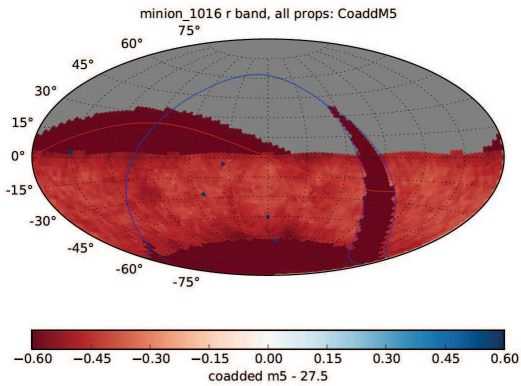} 
 \caption{Current observing program for LSST illustrated in an Aitoff projection of equatorial coordinates (Celestial poles at $\pm 90^\circ$); the Galactic plane is shown as a line bifurcating a dark curved region (the Galactic avoidance zone) to be visited only rarely in the course of the mission. The scale bar corresponds to the 5$\sigma$ $r$ magnitude depth $-27.5$ mag, at the end of a 10 year mission.  The impact of this nominal  program corresponds to almost nill cadence and a serious hit to astrometric programs in the Galactic plane.  The nominal baseline cadence illustrated here gives 30 exposures in all six filters to fields in the Galactic ``zone of avoidance'' that are to be completed in the first 200 days of the LSST WFD survey.  However, it may be surprisingly cost effective to extend  the baseline WFD survey to include the plane;  a final survey depth only 0.04 mag brighter would result and this could be repaired by allocating e.g. 90\% of the time to the WFD survey. ({ \tt astro\_lsst\_01\_1004} Figure 2, LSST Observing Strategy v1.0; Marshall et al. 2017, p.47)}
\label{fig3}
\end{center}
\end{figure}

The fundamental mission of LSST can be summed up as ``deep, fast, wide'' and more analogous to a satellite in operations, rather than a ground based telescope.  While not the largest telescope in aperture (6.7m effective) LSST will surpass any other facility by an order of magnitude in \'etendue, A$*\Omega=319 {\rm m^2 deg^2}$, the product of effective aperture in m$^2$ and field of view (9.6 deg$^2$; Fig \ref{fig2}).   The baseline observational plan (Marshall et al. 2017; LSST Observing Strategy White paper) has LSST taking pairs of 15 sec exposure ``visits'' with each field visited twice per night and the full sky done every 3 nights.  LSST will cover 18,000 deg$^2$ of high Galactic latitude sky with airmass $X<1.4$ with 825 visits per point, over 6 LSST filters, over the 10 year mission.   For a single visit this translates into median depth of $(ugrizy)=(23.9, 25.0, 24.7, 24.0, 23.3,22.1)$ with median seeing 0.81 arcsec.   The final depth is slated to achieve $(ugrizy)=(26.3, 27.5, 27.7, 27.0, 26.2, 24.9)$.  The baseline mission requires 90\% of the observing time and presently, the Galactic plane is seriously shortchanged due to the anticipated crowding (Figure \ref{fig3}).   To sum up, LSST is an optical/near IR survey of roughly half the sky in $ugrizy$ bands to $r\sim 27.5 = 36$ nJy, based on 825 visits over 10 years.   Per night, there will be $\sim 10$ million time domain events per night, detected and transmitted to event distribution networks within 60 sec of observation, a catalog of orbits for $\sim 6$ million bodies in the Solar System, a catalog of $37$ billion objects (20B galaxies, 17B stars, $\sim 7$ trillion observations (``sources''), and $\sim$ 30 trillion measurements (``forced sources''), produced annually, accessible through online databases, and deep, coadded images.  The services and computing resources will be available at Data Access Centers to enable user-specified custom processing and analysis; access to these will be for LSST members (https://www.lsst.org/about/dm/data-products).  

What happens with the remaining $\sim 10\%$ of observing time?  This is where the community has considerable input, through submission of programs for ``deep drilling'' fields and presumably, superior cadence in the Galactic plane.  Also it is noteworthy that all LSST observations will be tagged for image quality, so it is possible to build a dataset for e.g. the Galactic bulge using only images with the best seeing, potentially enabling one to somewhat beat the image crowding problem.   However, justification for Galactic plane surveys will face some challenges as facilities like PANSTARRS and especially KMTnet (a 3$\times$1.6 m telescope array funded by Korea; A$*\Omega=6$) aimed at microlensing planet surveys) have growing impact; KMTnet will have BVRI filters.  However, none will have the uniformity, data quality, data access, and software tools of LSST, consequently the Galactic science community should seriously focus on LSST as a {\it fundamental} long term opportunity.   A tool (OpSim) allows investigators to build observing programs and to test strategies that differ from the baseline observing cadence; its function is similar to that of APT for HST planning (Delgado et al. 2014).  The {\it present} weak coverage of the baseline cadence in the Galactic plane is shown in Figure \ref{fig3}; the poor cadence is a more serious issue than brighter limiting magnitude, as it will mean that transient and variable phenomena will be poorly studied in the plane if it is not remedied. 

\subsection{The case for extending ``wide, fast, deep'' to the Milky Way disk/bulge}
 While this review does not consider the Milky Way transient science case in depth, variability offers another  ``dimension'' in addition to astrometry and photometric metallicities in the exploration of the Milky Way.  Obviously, RR Lyrae and Cepheids are powerful distance indicators that can also map structure in the disk, plane, and even nearby galaxies.  Thanks to their high luminosity, miras can reach to great distances although with lower precision (see Catchpole et al. 2016).  Marshall et al. (2017) identify four additional science programs that might benefit if the plane is included: (1) Quantifying the large quiescent compact binary population via variability; these are massive black hole binaries with presently low X-ray luminosity.  (2) new insights into novae and the route to Type I SNe; LSST could improve drastically on the paltry rate at which novae are discovered (15 per year) and perhaps discover a host of new nova-like variables and stellar transients. (3) Discovering the next Galactic supernova; and (4) measuring population parameters of planets outside the snow line, using microlensing (Gould 2013).    This latter LSST science case would require very high cadence followup of candidates using 1-2 m class telescopes (KMTnet?) around the world.   The program seeks to return the statistics of planets moderately distant from their host star, poorly probed by both Doppler and transit techniques, yet not distant enough to yield to direct high contrast imaging surveys.    The bulge hosts $\sim 2 \times 10^{10} L_\odot $ stars, not only an exceptional laboratory for stellar transient phenomena, but the superb astrometry will enable proper motion separation along the lines of Kuijken \& Rich (2003) and Clarkson et al. 2008 over the whole of the bulge.  While the science goals of the Blanco DECam Bulge Survey, described below, include searching for a widespread intermediate age population claimed to be present in the bulge \cite{bensby17}, the LSST program offers perhaps the best opportunity to measure the bulge star formation history over the entire bulge.  

\section{Astrometry}

As illustrated in Figure 4, LSST will produce almost Gaia quality astrometry, but $\sim 3-4$ mag fainter.
Although the focal plane will present challenges (Bernstein et al. 2017a; see also Pier et al. 2003 for SDSS astrometry) perhaps the most impressive potential for LSST will be in its astrometric grasp.  The use of the Gaia astrometry to establish an astrometric grid may be a substantial improvement,
however there are significant issues connected with the atmosphere and secular change in Gaia positions e.g. Gaia catalog position, precession, nutation, aberration, Gaia proper motion, Gaia parallax, diurnal aberration, generic refraction, differential chromatic refraction based on $g,r,i,z,y$, spherical to tangent plane, and all optical corrections e.g. pincushion distortion etc. (Monet 2017; Bernstein et al. 2017a).  In the case of LSST, one may hope that the focal plane would be stable over long periods, however the design is modular and calls for it being possible to exchange out failed CCDs via modules; such an event might well spur the need for a new calibration series that should include astrometric calibration.  Also, the focal plane may change over time due to material fatigue.  An important goal of LSST is to extend in time the Gaia catalog past the end of the Gaia mission, and possibly extend Gaia astrometry to significantly fainter magnitudes.   In addition to all of the other gains of observing at the lowest possible airmass, observing on the meridian will be important if LSST is to produce the best astrometry.   The Galactic plane will host many high proper motion stars and the $izy$ photometric bandpasses will excel for detection and characterization of them, but multiple visits are mandatory to properly flag them as high PM stars and not simply transients.  {\it Developing and sustaining excellent astrometry for Galactic science will require multiple visits to the Galactic plane; the transient/variability and astrometric science cases both require that the Galactic plane not be shortchanged with respect to cadence.}   One also might seriously consider testing innovative approaches during commissioning time.  The use of rapid scanning techniques (Riess et al. 2014; Casertano et al. 2016) enables HST to gain a factor of 10 increase in astrometric precision (perpendicular to the scan) compared to a conventional centroid.  Might a scanning approach be similarly used to achieve super-precision for LSST centroiding and mapping of field distortion in the very stable LSST focal plane?  A twilight scan program in the Y band might enable a novel parallax program for nearby faint red dwarfs.  
\begin{figure}
\begin{center}
 \includegraphics[width=6.0 in]{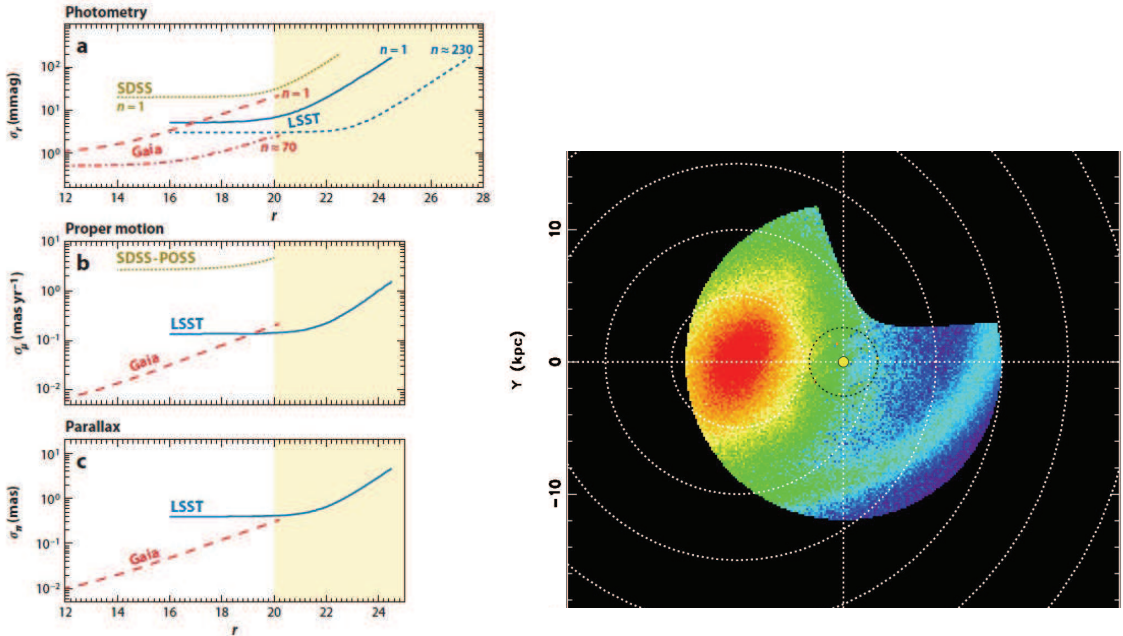} 
 \caption{ (Left): A comparison of (a) photometric errors, (b) proper-motion, and (c) parallax errors for the Sloan Digital sky survey (dotted line), Gaia (dashed line) and LSST (solid line) as a function of apparent magnitude for a G2V star.   The solid LSST curves show single visit accuracy; dashed LSST curve shows 230 visits in the $r$ band; LSST has the potential to extend Gaia astrometric precision to fainter magnitudes (Ivezi\'{c}, Beers, \& Juri\'{c} 2012; fig. 21; reproduced here by permission).  (Right) Approximate extent to which photometric metallicities and distances can be derived for dwarfs based on end of mission precision by applying methods of Ivezi\'c 2008 to the LSST dataset (source: LSST science book).}
   \label{fig4}
\end{center}
\end{figure}

\begin{figure}
\begin{center}
\includegraphics[width=3.8 in]{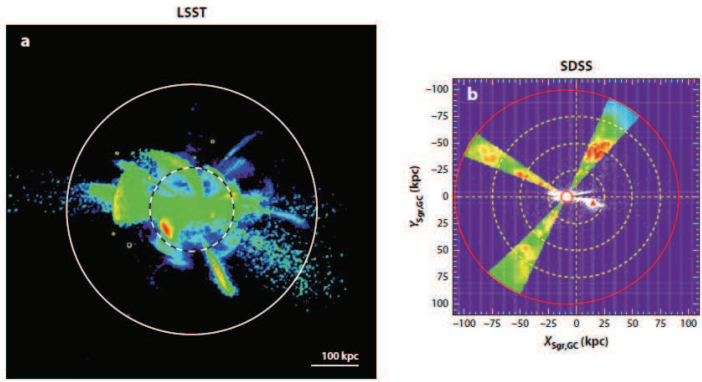} 
 \caption{Simulation of (a) the outer regions of the Milky Way compared to (b) the current state-of-the-art data shown on the same spatial scale. The data panel (b) presents the number density multiplied by the cube of the Galactocentric radius (logarithmic scale with dynamic range of 1,000, from blue to red ), for $\sim$ 1,000 Sloan Digital Sky Survey (SDSS) RR Lyrae stars within $10^\circ$ of the Sgr dwarf tidal stream plane (Ivezi\'{c} et al. 2004). The same color coding is used to visualize the stellar number density for a Milky Way type galaxy simulation (a) from Bullock \& Johnston (2005), shown on the left. Set within a CDM merger history, these simulations track the accretion and disruption of hundreds of dark matter halos into Milky Way size halos. With the Large Synoptic Survey Telescope (LSST),RR Lyrae stars will be found beyond the presumed Milky Way tidal radius ($\sim$300 kpc, white circle), and the much more numerous main-sequence stars will trace the structure significantly beyond the  $\sim100$ kpc (smaller black-and-white dashed circle) The latter distance range can presently only be accessed using  RR Lyrae and giants. Original color figure (Ivezi\'{c}, Beers, \& Juri\'{c} 2012; fig. 22; reproduced here by permission).}

\label{fig5}
\end{center}
\end{figure}

It is also important to emphasize that SDSS had essentially one epoch of imaging, but LSST will run for 10 years or longer, and will enjoy the synergy with Gaia; proper motions improve with time, so the power of the LSST astrometric program will grow over time.  The return on such investment will be that astrometry alone can be used for important discoveries, such as the identification of streams by common proper motion to great distances (tens of kpc) and possibly, Local Group proper motion studies that benefit from the use of thousands or even millions of stars (e.g. internal motions of the Magellanic clouds and the Sgr dwarf spheroidal galaxy/stream).    As emphasized in Ivezi\'{c} et al. (2012; fig 4),  the proper motion measurements of LSST will be comparable to those of Gaia at its faint limit $(r\sim 20).$   However, LSST carries Gaia's faint end proper motion accuracy  to $r=21.5 $ (Figure \ref{fig4} ).   In practical terms, proper motions of 0.2 mas/yr at r=21 correspond to roughly 10 km/sec  precision at 10 kpc (one sorts the bulge from the disk as easily as Kuijken \& Rich (2002) or Clarkson et al. 2008 did with HST, but over the whole bulge.  At r=24 the precision is 1.0 mas/yr or 60 km/sec precision (5 yr) at 60 kpc, possibly enough to sort out streams and high velocity halo stars.   In cases where streams and structures in the halo are not coherent in space or composition, they may still be detected using common proper motion.  The gains of LSST pushing fainter than Gaia by up to 7 mag are manifold.  Not only can structures traced by RR Lyrae be detected to 400 kpc (Figure \ref{fig5}), but the main sequence stars (100 times as abundant as red giants) in the ultrafaint dwarf galaxy population become accessible with depth near 27-28 mag of HST observations.  This translates into higher S/N detections of ultrafaint dwarf galaxies and streams, and intergalactic red giants being detectable to 1 Mpc.   One should think of stellar populations as a kind of iceberg; the red giants and RR Lyraes with $10^8$ yr lifetimes are the proverbial ``tip'' of the iceberg, but their numbers are set by the lifetimes of evolved stars.  Most of the ``iceberg" is in the long lived core hydrogen burning stellar population that by and large is fainter than the turnoff (roughly 1 $L_\odot$).  The reach of LSST, 3 mag past SDSS, will grasp the MSTO at $M_V\sim +4$ to distance moduli of 23 (200 kpc), RR Lyrae detections and proper motions (10 year mission) possibly to 400 kpc (for a taste of what is feasible, see the spectacular mapping of RR Lyrae in the Sgr dwarf to 80 kpc and 3\% distance precision from Panstarrs data; Hernitschek et al. 2017)   If detected stream members can be increased dramatically, we may detect with confidence stream gaps or disturbances arising from interaction with dark matter cores (Carlberg 2013). 

Another promising tool deriving from the LSST dataset will be the use of the {\it reduced proper motion}, presently most familiar to the astrometry community.   The quantity $H_m = m + 5 \log \mu + 5$ is the reduced proper motion, where $\mu$ is the proper motion in units of arsec yr$^{-1}$ and m is a magnitude bandpass.  Paired with a wide color baseline and photometric metallicities, the reduced proper motion may uncover coherent structures at great distances, and can also sort out nearby high proper motion stars in a proxy color-magnitude diagram, especially the subdwarfs vs white dwarfs (see e.g. Majewski 1999, Lepine, Rich, \& Shara (2007), and Carlin et al. 2012; but see a cautionary note concerning disk/halo separation in Appendix B of Sesar et al. 2008).  With a high precision measurement of $\mu$ being available for every star, along with 6 color photometry, a calibrated color baseline that extends from UV to 1$\mu \rm m$, and reddening-free indices, LSST will be a more powerful dataset than many realize, even without a spectroscopic counterpart.  {Indeed, for Milky Way studies, the photometric metallicities + proper motions translate into potential ``spectroscopy" for billions of stars reaching to 100 kpc.}  Although the VVV survey is a pure infrared survey, the $y$ filter of LSST will have generally superior image quality and calibration and will reach 22.1 (single epoch) and 24.9 coadded.  Recall that subdwarfs and cool white dwarfs will be faint and problematic for Gaia (the g=21 limit is essentially that of of the Palomar Observatory Sky Survey) whereas the infrared sensitivity of LSST can reach these populations at nearly 1 kpc for a dwarf with $M_V=+15$.   Finally the six-color photometry will enable not only stellar classification, but also boost star/galaxy separation even for unresolved galaxies, via photometric redshifts (galaxies) and spectral type classification (stars). 

At this meeting we have heard about the 4MOST spectroscopic survey facility (both medium and high resolution spectroscopy of $10^5-10^6$ stars with a 4m telescope; larger spectroscopic survey telescopes in the 8-10 m class may be on the horizon.)  It is noteworthy that 4MOST will commence operations early in the LSST mission.  LSST photometry has the likelihood or yielding useful photometric metallicities but will not give any constraint on detailed abundances.  LSST may produce a virtually complete map of the Milky Way halo; individual red giants will be classifiable to 1 Mpc and the reach extends well into the Local Group, but high resolution spectroscopy will be required to elucidate the rich chemical evolution history of these structures.  

\section{The Blanco DECam Bulge Survey:  An LSST Milky Way Pathfinder Project}

\begin{figure}
\begin{center}
 \includegraphics[width=3.0 in]{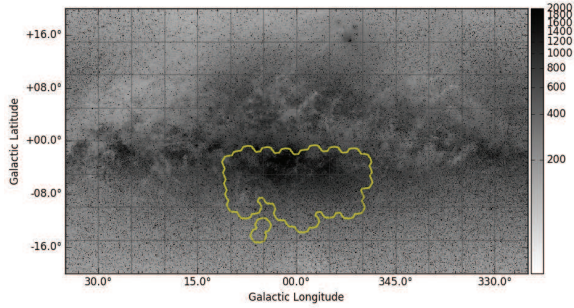} 
 \vspace*{-0.2 cm}
 \caption{ The Blanco DECam bulge survey is a $\sim 200$ deg$^2$ survey of the Galactic bulge (outlined in the scalloped line) using DECam at the Blanco 4m telescope; we have one billion sources detected in $ugrizy$; figure courtesy of BDBS co-I W. Clarkson.}.
   \label{fig6}
\end{center}
\end{figure}

\begin{figure}
\begin{center}
 \includegraphics[width=4.0 in]{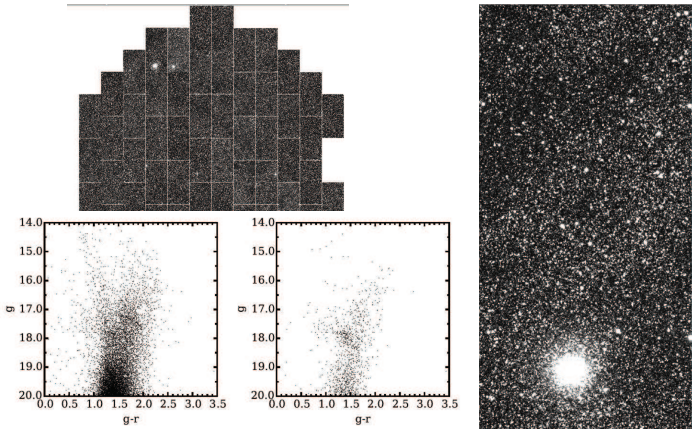} 
 \vspace*{-0.2 cm}
 \caption{ (Left): upper portion illustrates the top of an r band DECam image of the field including the globular cluster NGC 6569.  Color-magnitude diagrams derived using daophot below refer to the individual CCD frame on the right. They are single 75 sec $g, r$ frames.  The color-magnitude diagram on the left is that of the field population, while that on the right is that of the NGC 6569 globular cluster (Johnson et al. 2017).  The red clump is clearly visible.  The entire 30 TB dataset has been reduced and is in the final stages of catalog merging. }.
   \label{fig7}
\end{center}
\end{figure}
\vspace*{-0.1 cm}
We have concluded the first epoch of observations of the Blanco DECam Image survey using $ugrizy$ imaging with the Dark Energy Camera.  
As the Galactic bulge is investigated using spectroscopic and photometric surveys, its apparent complexity grows. The evidence for the overall bar structure is well established (see e.g. Rich 2013).  An X shaped bulge/bar structure \cite{mcwzoc} has been mapped using the VVV infrared survey \cite{wegg13}.  Each new study suggest the population is more complex as a function of age, metallicity, and spatial distribution.   It has long been known that there is a strong primary correlation between higher metallicity and decreasing velocity dispersion, but new studies argue for multiple spatial components and multimodality in abundance distributions (see e.g. \cite{zoc17} and Bensby et al. 2017).  The RR Lyrae stars (Kunder et al. 2016) appear to lack rotation; the short period (older?) miras do not follow the bar distribution, while the long period (intermediate age?) miras follow the bar (Catchpole et al. 2016).  One approach is to address the problem along the lines of \cite{ivezic08} and to use photometric metallicities to map the bulge structure using millions of stars.  Even in crowded fields like Baade's Window, our photometry reaches fainter than the main sequence turnoff.  We can investigate assertions of a substantial intermediate age population over the contiguous bulge field of regard.  A further question arises concerning the nature of the bulge blue horizontal branch; there is the opportunity to match groundbased $u$ band data imaging from the near-UV channel of the Galaxy Evolution Explorer.  Our overarching goal is to map the structure of the bulge as a function of age and metallicity, and to explore the spatial distribution of populations such as blue horizontal branch stars and metal rich stars.   C. Johnson (CfA) has reduced the full 30 TB dataset using a parallelized version of daophot; the reductions are being done using the cluster at the Pervasive Technology Institute (PTI) in collaboration with Scott Michael and Mike Young.   Our goal will be to have a catalog released to the community 12 months after we have certified a final dataset.  We also aim to have a postage stamp image server operating along with the catalog release.

{ \sl Acknowledgements}:  I wish to extend special thanks to Z. Ivezic and N. Kallivalyalil for suggestions and a critical reading of the manuscript, and W. Clarkson for sharing a presentation and useful suggestions.  I acknowledge support from grant AST-1413277 from the NSF.  The author acknowledges the profound contributions of Branimir Sesar, who has left the field; his presence in the LSST era will be sorely missed.

\end{document}